\documentclass[a4paper,11pt]{article}
\usepackage{pos}
\usepackage{float}
\usepackage{amsmath}
\usepackage{mathrsfs}
\usepackage{graphicx}
\usepackage[export]{adjustbox}
\usepackage{lineno}
\usepackage{caption}
\usepackage{subcaption}

\title{Measuring the Neutrino Cross Section Using 8 years of Upgoing Muon Neutrinos Observed with IceCube}
\ShortTitle{Neutrino Cross Section}

\author{The IceCube Collaboration \\{\normalsize \normalfont(a complete list of authors can be found at the end of the proceedings)}}

\emailAdd{srobertson@icecube.wisc.edu}

\abstract{The IceCube Neutrino Observatory detects neutrinos at energies orders of magnitude higher than those available to current accelerators.  Above 40 TeV, neutrinos traveling through the Earth will be absorbed as they interact via charged current interactions with nuclei, creating a deficit of Earth-crossing neutrinos detected at IceCube. The previous published results showed the cross section to be consistent with Standard Model predictions for 1 year of IceCube data. We present a new analysis that uses 8 years of IceCube data to fit the $\nu_\mu$ absorption in the Earth, with statistics an order of magnitude better than previous analyses, and with an improved treatment of systematic uncertainties.
It will measure the cross section in three energy bins that span the range 1 TeV to 100 PeV.  We will present Monte Carlo studies that demonstrate its sensitivity.

\vspace{4mm}
{\bfseries Corresponding author:}
Sally Robertson$^{1}$ $^{2*}$\\
{$^{1}$ \itshape University of California Berkeley}\\
{$^{2}$ \itshape Lawrence Berkeley National Laboratory}\\[4mm]
$^*$ Presenter

\FullConference{37$^{\rm{th}}$ International Cosmic Ray Conference (ICRC 2021)\\
		July 12th -- 23rd, 2021\\
		Online -- Berlin, Germany}

}

\begin{document}
\maketitle

\section{Introduction}

Neutrinos are weakly interacting particles
which interact through charged current and neutral current interactions.
The cross section of neutrino interactions has only been measured in neutrino beams at accelerator experiments up to $400\, \mathrm{GeV}$.
In contrast, IceCube detects neutrinos from about $10\, \mathrm{GeV}$ to above $1$~PeV,  orders of magnitude higher than accelerator experiments. 

The IceCube neutrino detector is subject to a continuous diffuse flux of neutrinos coming through the Earth which are mostly produced in cosmic ray air showers, with a fraction coming from astrophysical sources \cite{6yrDiffuse}. IceCube uses Digital Optical Modules (DOMs) arranged in a grid to detect Cherenkov light from the charged secondary particles produced in neutrino interactions \cite{IceCubeInst}.

The neutrino-nucleon cross-section rises with energy, so that at energies above about 40 TeV the Earth becomes opaque to neutrinos. IceCube can be used to find the flux of neutrinos as a function of energy and zenith angle. The flux of neutrinos at the horizon will be different than the flux coming vertically through the Earth as neutrinos are absorbed.  To measure the cross section the observed flux, as a function of energy and zenith angle, is compared with models that use different absorption cross-sections, and the best-fit cross-section is determined. The analysis accounts for the density profile of the Earth's interior, based on the PREM model \cite{prem}. 

Measuring the neutrino cross section has potential to detect Beyond Standard Model (BSM) processes which will occur at high neutrino energies and appear as increased absorption \cite{KleinChapter}. IceCube also has potential of measuring nuclear processes such as nuclear shadowing.  For neutrinos travelling through the Earth's core shadowing can alter the cross section by $3\%- 4 \%$ at energies above $100$ TeV \cite{nshadowing}. 
Other interactions, such as diffractive Coulomb induced events like W-boson and trident production \cite{1998Seckel, trident} can increase the total cross-section.

\section{Previous Measurements}
Currently there are three cross section measurements above TeV energies: two previous IceCube cross section measurements that use through going and starting tracks respectively \cite{xsec, HESExsec}, and a third that used publicly available IceCube cascade events \cite{xsecBust}. 
IceCube's first multi-TeV neutrino cross section measurement used one year of through going muon data collected in 2010 when IceCube was still under construction and consisted of 79 strings out of its final 86. That analysis used
10,784 upgoing muon neutrino events, and measured the total neutrino cross section as $R=1.30_{-0.19}^{+0.21}(stat.)_{-0.43}^{+0.39}(syst.)$
times the Standard Model prediction
in the energy range from 6.3 TeV to 980 TeV\cite{xsec}.

\begin{figure}[htp]
	\centering
	\includegraphics[width=0.9\textwidth]{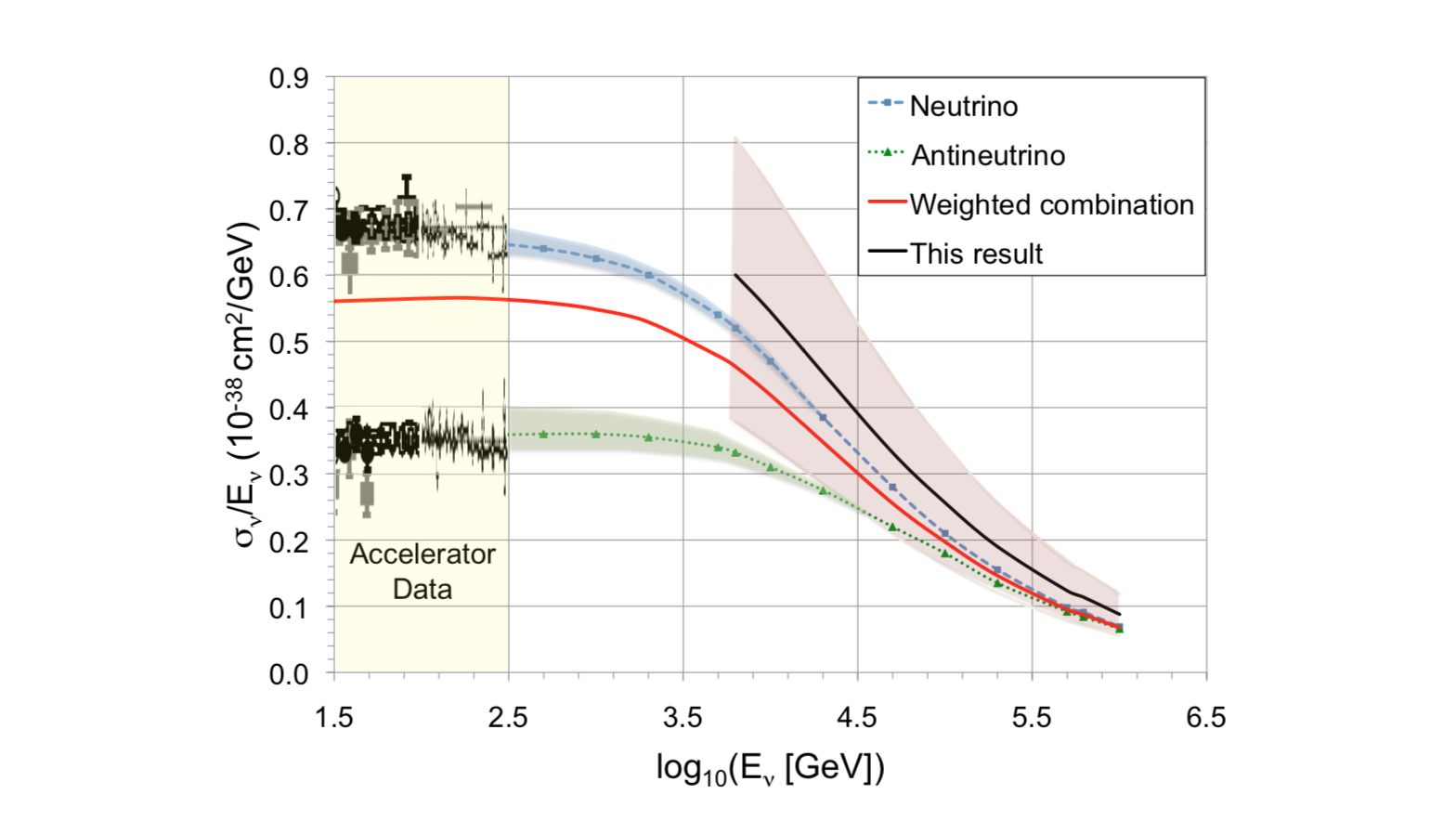}
	\caption{Compilation of neutrino charged current cross section measurements divided by neutrino energy, from accelerator experiments \cite{particledata}, theoretical prediction from \cite{xsecCooper} and IceCube's current result \cite{xsec}. Figure from \cite{xsec}.  \label{result}}
\end{figure}

IceCube also has performed a multiflavour cross section analysis using High Energy Starting Events (HESE) \cite{HESExsec}. The HESE sample contains events from 60 TeV to 10 PeV, an order of magnitude higher than the one year upgoing measurement. 
The independent analysis used six years of public IceCube data  to study cascades with energies between 18 TeV and 2 PeV \cite{xsecBust}. Both analyses were found to be consistent with the standard model cross section. 

\section{Extended Cross Section  Analysis}

The analysis presented here uses 8 years of upgoing neutrino data from 2011-2018 \cite{ICRCJoeran}, comprising approximately 300,000 muon neutrino events. This larger data sample will result in the reduction of statistical uncertainty by a factor of 10 compared to the previous analysis, at the same time extending the energy range of the measurement down to $ 10^{3}$ GeV and up to $10^{8}$ GeV.  The analysis will be done in three energy bins, with the cross section determined independently in each bin. Here, the energy bins cover energy ranges $10^{3} - 10^{4}$ GeV (bin $x_1$), $10^{4}- 10^{6}$ GeV (bin $x_2$) and $10^{6}-10^{8}$ GeV (bin $x_3$)).

\subsection{Extended Cross Section Analysis Method}
This analysis uses an updated calculation for the neutrino transmission through the Earth, which uses the nuSQuIDS library.
This program relies on SQuIDs \cite{squid} which uses the evolution of quantum mechanical ensembles to calculate the propagation of a neutrino, taking into account interactions and oscillations. The nuSQuIDS library was adjusted to introduce the variable $R$, which is the multiple of the standard model cross section applied to both charged current and neutral current interactions. For neutral current interactions the outgoing lepton will be a neutrino with reduced energy, thus the change in cross section will change in the energy distribution. For charged current interactions the neutrino is lost in the interaction. 

\begin{figure}[ht]
\begin{center}
\includegraphics[width = 0.49\textwidth]{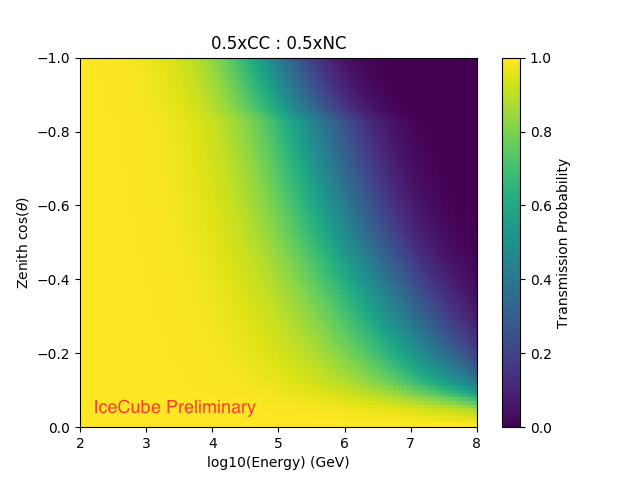}
\includegraphics[width = 0.49\textwidth]{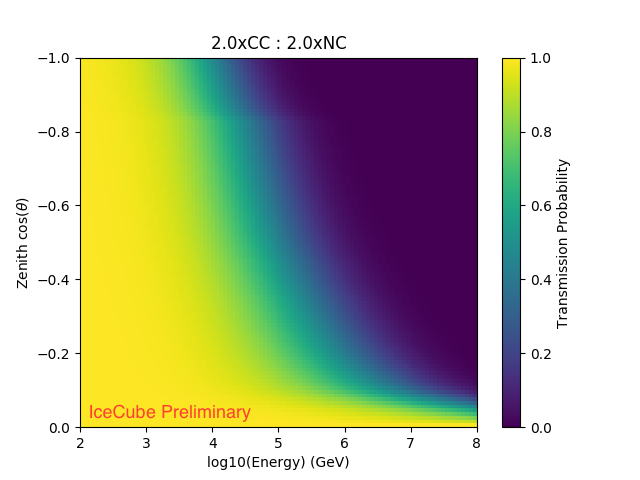}
\end{center}
\caption[]{Transmission probability of neutrinos traveling through the Earth with modifications to the standard model cross section by a multiplicative factor R. (\emph{left} R=0.5, \emph{right} R=2)
}
\label{fig:tp}
\end{figure}

\begin{figure}[ht]
\begin{center}
	\includegraphics[width = 0.49\textwidth]{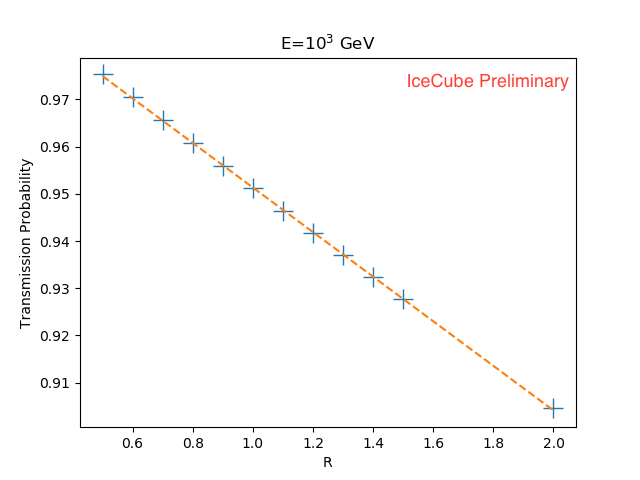}
	\includegraphics[width = 0.49\textwidth]{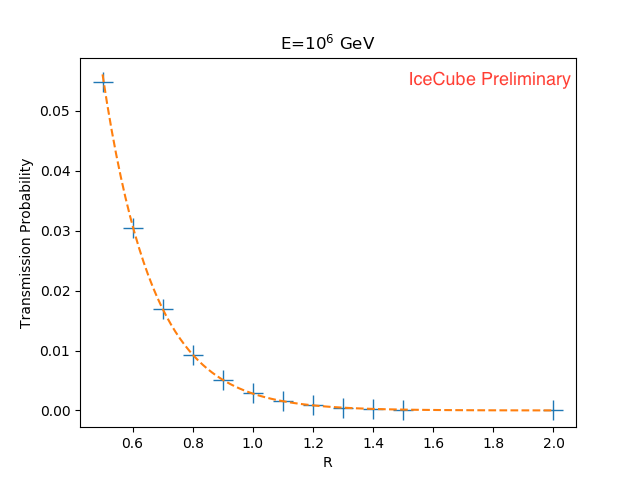}
\end{center}
\caption[]{Transmission probability for a vertical neutrino ($\cos\theta $=-1.0) for (\emph{left}) E=$10^3$ GeV and (\emph{right})  E=$10^6$ GeV. The line is a linear fit to the points on the left and an exponential fit on the right.
For given values of $R$ and energy, the highest sensitivity to $R$ is at path lengths corresponding to moderate absorption.
}
\label{fig:inter}
\end{figure}

Transmission probabilities were determined by adjusting the cross section by multiples of the standard model with $R$ being the same for both neutral current and charged current interactions. 
Figure \ref{fig:tp} shows the transmission probabilities for two values of $R,$ vs. zenith angle and energy, illustrating the change in transmission for two different cross section multipliers (zenith angle $\cos\theta=-1.0$ is vertically upward going). For each Monte Carlo event a function was created to interpolate the transmission probability vs. $R$.
Figure \ref{fig:inter} shows an example of the transmission probability as a
function of $R$ for a vertically traveling neutrino. The red dashed line
is a fit to the discrete values of $R$ shown. For low energy events a linear fit was used while above energies of $10^{5}$ GeV, an exponential function was used.

This analysis relies on the same fitting method as the IceCube diffuse muon neutrino analysis \cite{6yrDiffuse,ICRCJoeran}, which performs a Poisson likelihood fit with priors for the muon neutrino flux to the atmospheric and astrophysical predictions. 
The cross section is introduced to the fit as a new parameter which applies a re-weighting to the flux.
The re-weight is
calculated from the interpolation of the transmission probability across $R$. Figure \ref{fig:inter} shows an example but each MC event has its own interpolation calculated for its energy and zenith angle. 

The systematic uncertainties follow from the IceCube analysis described in Ref. \cite{ICRCJoeran}, which included new parameters for the ice model and the use of an updated atmospheric neutrino flux. The systematic uncertainties on the ice model were included in the previous analysis to account for the uncertainties in the absorption and scattering of the Cherenkov light. A new systematic is now included for the larger scattering in the refrozen ice around each DOM. The atmospheric fluxes now use the MCEq package \cite{MCEq} which uses the SIBYLL23c model \cite{SIBYLL}. The fit now also uses the Barr parameters for the uncertainties on hadronic production in cosmic ray air showers as calculated in \cite{barr}.

\subsection{Monte Carlo Tests for the Extended Analysis}

\begin{figure}[ht]
    \begin{center}
    \begin{subfigure}[b]{0.45\textwidth}
        \includegraphics[width=\textwidth]{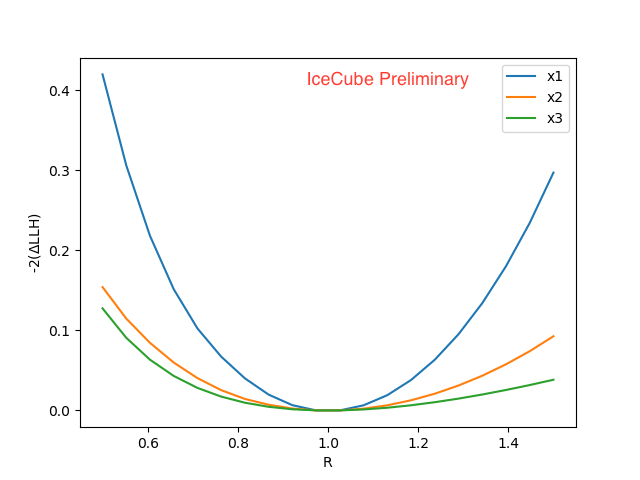}
        \caption{}
        \label{fig:asimov1d}
    \end{subfigure}
    \begin{subfigure}[b]{0.45\textwidth}
        \includegraphics[width=\textwidth]{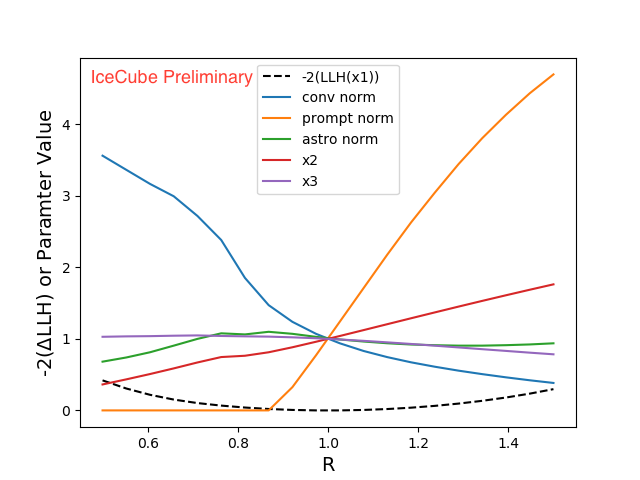}
        \caption{}
        \label{fig:systematic1d}
     \end{subfigure}
    \end{center}
    \caption[]{ The Asimov likelihood fit across modification of cross section $R$, for (a) each scan is a separate fit for the given energy bin. 
    In (b) the likelihood fit is for a scan over the energy bin $x1$ showing how the nuisance parameters change with the fit. The parameters with the most effect on the cross section are the conventional and prompt normalizations. 
    }
    \label{fig:asimov}
\end{figure}

To check the consistency of the analysis, Asimov fits were performed on
the Monte Carlo sample instead of using pseudo data \cite{asimov}. 
The Asimov likelihood scan for each energy bin is shown in Fig. \ref{fig:asimov1d}. The lowest energy bin, $x_1$, has the narrowest likelihood due to the large number of events in the energy bin. The energy ranges of the higher energy bins, $x_2$ and $x_3$, have been chosen to have similar likelihood widths. 

In the Monte Carlo fits changes to the cross section are compensated by changes to the flux normalization for the conventional, prompt and astrophysical fluxes, see Fig. \ref{fig:systematic1d}. 
In some cases the normalizations can fit to values beyond their known physical limits. 
The final analysis will use priors in the fit to constrain the atmospheric and astrophysical fluxes. The priors are listed in Table \ref{tab:priors}. The conventional and prompt normalization are based on MCEq using SIBYLL23c \cite{SIBYLL}. 
The astrophysical normalization prior is based on the HESE result, which is independent of the $\nu_\mu$ results. 
The widths of the prior are kept large so as to cover the range of all current results. 

\begin{table}[ht]
    \centering
    \begin{tabular}{c | c}
        \hline
        Parameter & Prior/Width (Range)\\
        \hline
        Astrophysical Norm & $2.1 \pm 0.6$ (0,$\infty$) \\
        Convention Norm & $1 \pm 0.25$ (0,$\infty$)\\
        Prompt Norm & $1 \pm 2$ (0,$\infty$)\\
        \hline
    \end{tabular}
    \caption{The Gaussian priors on the flux normalization for the likelihood fit for the final analysis. Additional priors are on the systematic parameters.}
    \label{tab:priors}
\end{table}

\section{Conclusions}

The extended cross section analysis presented here is based on a
previous IceCube analysis but improvements have been implemented
both in the tools used and in the treatment of systematic uncertainties.
It also extends the lifetime of the data sample analyzed to 8 years.
The analysis is in the finalizing stages and moving towards unblinding of the whole data set. 
The results will extend the current IceCube
neutrino-nucleon cross section measurement to the energy range $ 10^{3}- 10^{8}$ GeV, with measurements made for multiple energy bins allowing for the possibility of detecting variations in cross section with neutrino energy.



\bibliographystyle{ICRC}
\bibliography{}

\clearpage
\section*{Full Author List: IceCube Collaboration}




\scriptsize
\noindent
R. Abbasi$^{17}$,
M. Ackermann$^{59}$,
J. Adams$^{18}$,
J. A. Aguilar$^{12}$,
M. Ahlers$^{22}$,
M. Ahrens$^{50}$,
C. Alispach$^{28}$,
A. A. Alves Jr.$^{31}$,
N. M. Amin$^{42}$,
R. An$^{14}$
% ...
\\

\noindent
% ...
$^{12}$ Universit{\'e} Libre de Bruxelles, Science Faculty CP230, B-1050 Brussels, Belgium \\
% ...
$^{14}$ Department of Physics and Laboratory for Particle Physics and Cosmology, Harvard University, Cambridge, MA 02138, USA \\
% ...
$^{17}$ Department of Physics, Loyola University Chicago, Chicago, IL 60660, USA \\
$^{18}$ Dept. of Physics and Astronomy, University of Canterbury, Private Bag 4800, Christchurch, New Zealand \\
% ...
$^{22}$ Niels Bohr Institute, University of Copenhagen, DK-2100 Copenhagen, Denmark \\
% ...
$^{28}$ D{\'e}partement de physique nucl{\'e}aire et corpusculaire, Universit{\'e} de Gen{\`e}ve, CH-1211 Gen{\`e}ve, Switzerland \\
% ...
$^{31}$ Karlsruhe Institute of Technology, Institute for Astroparticle Physics, D-76021 Karlsruhe, Germany  \\
% ...
$^{42}$ Bartol Research Institute and Dept. of Physics and Astronomy, University of Delaware, Newark, DE 19716, USA \\
% ...
$^{50}$ Oskar Klein Centre and Dept. of Physics, Stockholm University, SE-10691 Stockholm, Sweden \\
% ...
$^{59}$ DESY, D-15738 Zeuthen, Germany \\
% ...

\subsection*{Acknowledgements}

\noindent
USA {\textendash} U.S. National Science Foundation-Office of Polar Programs,
U.S. National Science Foundation-Physics Division,
U.S. National Science Foundation-EPSCoR,
Wisconsin Alumni Research Foundation,
Center for High Throughput Computing (CHTC) at the University of Wisconsin{\textendash}Madison,
Open Science Grid (OSG),
Extreme Science and Engineering Discovery Environment (XSEDE),
Frontera computing project at the Texas Advanced Computing Center,
U.S. Department of Energy-National Energy Research Scientific Computing Center,
Particle astrophysics research computing center at the University of Maryland,
Institute for Cyber-Enabled Research at Michigan State University,
and Astroparticle physics computational facility at Marquette University;
Belgium {\textendash} Funds for Scientific Research (FRS-FNRS and FWO),
FWO Odysseus and Big Science programmes,
and Belgian Federal Science Policy Office (Belspo);
Germany {\textendash} Bundesministerium f{\"u}r Bildung und Forschung (BMBF),
Deutsche Forschungsgemeinschaft (DFG),
Helmholtz Alliance for Astroparticle Physics (HAP),
Initiative and Networking Fund of the Helmholtz Association,
Deutsches Elektronen Synchrotron (DESY),
and High Performance Computing cluster of the RWTH Aachen;
Sweden {\textendash} Swedish Research Council,
Swedish Polar Research Secretariat,
Swedish National Infrastructure for Computing (SNIC),
and Knut and Alice Wallenberg Foundation;
Australia {\textendash} Australian Research Council;
Canada {\textendash} Natural Sciences and Engineering Research Council of Canada,
Calcul Qu{\'e}bec, Compute Ontario, Canada Foundation for Innovation, WestGrid, and Compute Canada;
Denmark {\textendash} Villum Fonden and Carlsberg Foundation;
New Zealand {\textendash} Marsden Fund;
Japan {\textendash} Japan Society for Promotion of Science (JSPS)
and Institute for Global Prominent Research (IGPR) of Chiba University;
Korea {\textendash} National Research Foundation of Korea (NRF);
Switzerland {\textendash} Swiss National Science Foundation (SNSF);
United Kingdom {\textendash} Department of Physics, University of Oxford.

\end{document}